\documentclass[conference]{IEEEtran}

\usepackage{cite}
\usepackage{amsmath,amssymb,amsfonts}
\usepackage{graphicx}
\usepackage{textcomp}
\usepackage{xcolor}
\usepackage{listings}
\usepackage{url}
\usepackage{booktabs}
\usepackage{array}
\usepackage{hyperref}
\usepackage{orcidlink}
\usepackage{tikz}
\usetikzlibrary{arrows.meta,decorations.pathreplacing,positioning}

\hypersetup{
  pdftitle={Balancing Workload Performance and Slurm Stress: Four Nextflow Deployment Strategies},
  pdfauthor={Nil Tianchen Mu, William Dizon, Glen Otero, and Torey Battelle},
  pdfkeywords={workflow management systems, Nextflow, HPC, Slurm, job scheduling, HyperQueue, Flux, benchmarking},
  colorlinks=true,
  linkcolor=blue,
  citecolor=blue,
  urlcolor=blue,
}

\begin{document}

\title{Balancing Workload Performance and Slurm Stress:\\
Four Nextflow Deployment Strategies}

\author{
\IEEEauthorblockN{Nil Tianchen Mu\IEEEauthorrefmark{1}\,\orcidlink{0000-0002-0271-0941},
William Dizon\IEEEauthorrefmark{1}\,\orcidlink{0009-0004-5656-9367},
Glen Otero\IEEEauthorrefmark{1}\,\orcidlink{0009-0007-7968-1433},
Torey Battelle\IEEEauthorrefmark{1}\,\orcidlink{0000-0003-0655-7355}}
\IEEEauthorblockA{\IEEEauthorrefmark{1}Research Computing, Arizona State University, Tempe, AZ, USA\\
Email: \{nilmu, wdizon, gotero3, tbattell\}@asu.edu}
}
\maketitle

\begin{abstract}
Wide Nextflow fan-outs on shared Slurm clusters can submit tens of thousands of
short tasks. Deployment settings route them through individual jobs, arrays, or
nested schedulers inside enclosing allocations. These settings determine
workflow turnaround and RPC volume, a shared cost that can degrade scheduler
responsiveness. Existing comparisons evaluate whole workflow systems, while
per-task queueing metrics cannot span architectures that dispatch inside
existing allocations.

We contribute a reproducible measurement protocol and benchmark harness. A
clean-start clock begins before backend startup or allocation requests, placing
architecturally different backends on a common time axis. Per-user Slurm
\texttt{sdiag} counters attribute request count as the primary \emph{RPC demand}
measure and controller processing time as sensitivity context, separate from
cluster-wide state. We apply the method to Slurm native dispatch, Slurm job
arrays, HyperQueue, and Flux on the shared ASU Phoenix production cluster and a
single-user Dev cluster. On Phoenix, every aggregation strategy improves both
objectives relative to native dispatch; Flux has the lowest RPC demand, while
HyperQueue's fastest median walltime is not stable across replicates. On Dev,
arrays and Flux improve walltime, while HyperQueue trades the lowest RPC demand
for the slowest completion. The method lets an HPC site compare deployment
strategies using both user-visible performance and scheduler impact, then select
the fastest strategy within its own RPC-demand limit.
\end{abstract}


\section{Introduction}
\label{sec:intro}

A standardized nf-core pipeline does not prescribe how it reaches Slurm.
Nextflow can submit individual jobs, form arrays, or delegate dispatch to
HyperQueue~\cite{beranek2024hyperqueue} or Flux~\cite{ahn2014flux} inside
enclosing allocations. This small configuration choice controls both turnaround
and RPC demand, yet is generally made by convention rather than measurement.

We measure workflow performance and Slurm control-plane demand for standardized
Nextflow~\cite{ditommaso2017nextflow} and nf-core~\cite{ewels2020nfcore}
pipelines on Slurm~\cite{yoo2003slurm}. The target class has a dominant fan-out
of short independent tasks; production whole-genome alignment can contain tens
of thousands of such tasks~\cite{mu2025optimizing}. The comparison focuses on
workflow turnaround and scheduler demand; reserved idle capacity and cluster
fragmentation are out of scope.

On a shared, Fairshare-scheduled cluster this fan-out has two distinct effects
that the WMS literature has largely separated. Every native task adds a
submission RPC and repeated status RPCs. Separately, the initial wave consumes
enough TRES usage to drive the association's Fairshare factor toward zero, after
which later tasks progress mainly through
backfill~\cite{mu2025optimizing,mu2026rmacc}.
Section~\ref{sec:tradeoff-fairshare} separates the two mechanisms. A deployment
can therefore improve user time while worsening RPC demand, or relieve
Slurm by paying for a larger enclosing allocation. Each choice also
changes acquisition delay, multi-node reach, accounting visibility, policy
enforcement, and failure isolation.

Which choice is right remains site- and workload-dependent;
Section~\ref{sec:positioning} positions this measurement question against prior
workflow, pilot-job, and scheduler work.

This paper contributes the measurement method for that question, and
characterizes the Nextflow-to-Slurm dispatch interface with it. Two obstacles
make the question hard to measure. Per-task queueing
metrics~\cite{feitelson1998metrics} cannot compare a Slurm task to a task
dispatched inside an allocation that already exists, because the two wait in
different places. A cluster-wide scheduler reading, in turn, cannot say which
co-tenant caused it, so it cannot price one deployment's demand. Each element of
the method removes one obstacle:

\begin{itemize}
\item \textbf{C1: a clean-start protocol.} A common clock opens before backend
initialization and before the first Slurm request, placing direct and
allocation-backed strategies on one walltime axis.

\item \textbf{C2: disciplined use of existing Slurm instrumentation.} Per-user
\texttt{sdiag} counters, a separate observer identity, boundary snapshots, and
an RPC-free settle tail attribute one deployment's requests and observed
controller processing time separately from cluster-wide Slurm state.

\item \textbf{C3: an empirical application.} We apply the method to four
deployment strategies on private and production Slurm clusters, exposing the
observed trade-off between workflow performance and attributable RPC demand.
\end{itemize}

\section{Background and Related Work}
\label{sec:background}

\subsection{Nextflow executors on Slurm}
\label{sec:executors}
Nextflow hands each process invocation to an \emph{executor} that places it on a
backend. We select four strategies spanning per-task dispatch, Slurm-native
aggregation, and nested schedulers
(Table~\ref{tab:backends})~\cite{nextflow_array_docs,beranek2024hyperqueue,ahn2014flux}.

\textbf{Terminology.} We call the long-running Nextflow driver the \emph{main
job}, and each Slurm job it submits per pipeline task a \emph{per-task job}. We
avoid ``child job'': at the Slurm level these are independent submissions, not
scheduler-tracked dependents. Every strategy submits one main job, so
Table~\ref{tab:backends} lists it beside the further Slurm objects each strategy
creates. For Flux, \(K\) denotes the fixed node count of its enclosing
allocation; Table~\ref{tab:factors} gives \(K\) for each cluster.

\begin{table*}[t]
\caption{Four multi-node deployment strategies. Their Slurm objects define the
dispatch, control, visibility, and TRES-accounting units.}
\label{tab:backends}
\centering
\footnotesize
\renewcommand{\arraystretch}{1.15}
\setlength{\tabcolsep}{3pt}
\begin{tabular}{@{}p{0.13\textwidth} p{0.13\textwidth} p{0.30\textwidth} p{0.13\textwidth} p{0.17\textwidth}@{}}
\toprule
\textbf{Backend} & \textbf{Executor} & \textbf{Slurm objects created} & \textbf{Dispatched by} & \textbf{Slurm-level control} \\
\midrule
Slurm native    & \texttt{slurm}                & main job $+$ one job per task            & Slurm        & per task \\
Slurm job array & \texttt{slurm}+\texttt{array} & main job $+$ one array job per batch     & Slurm        & per element \\
HyperQueue      & \texttt{hq}                   & main job $+$ HQ-requested worker allocations & HQ server & outer only \\
Flux            & \texttt{flux}                 & main job \emph{is} the $K$-node allocation, plus one \texttt{srun} step & Flux brokers & outer only \\
\bottomrule
\end{tabular}
\end{table*}

\subsection{Meta-schedulers on Slurm}
\label{sec:metaschedulers}
HyperQueue obtains Slurm worker allocations and dispatches tasks through an
internal server, whereas Flux starts a hierarchical broker inside an enclosing
Slurm allocation~\cite{beranek2024hyperqueue,ahn2014flux}. In both cases Slurm
accounts for enclosing resources but cannot control internal tasks separately;
aggregation changes accounting granularity rather than evading Fairshare.

\subsection{Positioning against existing benchmarking infrastructure}
\label{sec:positioning}
WfCommons~\cite{coleman2022wfcommons} supplies complementary synthetic-workflow
infrastructure; we use a real standardized pipeline so task and request
distributions are observed. ExaWorks~\cite{alsaadi2021exaworks} integrates
large-scale workflow technologies, while prior bioinformatics
comparisons~\cite{larsonneur2018evaluating,wratten2021review} compare whole WMS.

Dispatch inside a granted allocation is the pilot-job
pattern~\cite{turilli2018pilot}; Parsl~\cite{babuji2019parsl} and
Balsam~\cite{salim2019balsam} likewise aggregate task streams behind fewer batch
jobs.

Slurm architecture and guidance describe aggregation benefits and
message-, scheduling-, and lock-sensitive throughput
limits~\cite{jette2023slurm,slurm_arrays,slurm_high_throughput}; real-machine
simulator validation adds scheduler-level evidence~\cite{jokanovic2018simulator}.
They do not attribute controller demand to one workflow identity across dispatch
strategies under production co-tenancy. We measure that narrower trade-off.

\section{Measurement Method and Study Harness}
\label{sec:harness}

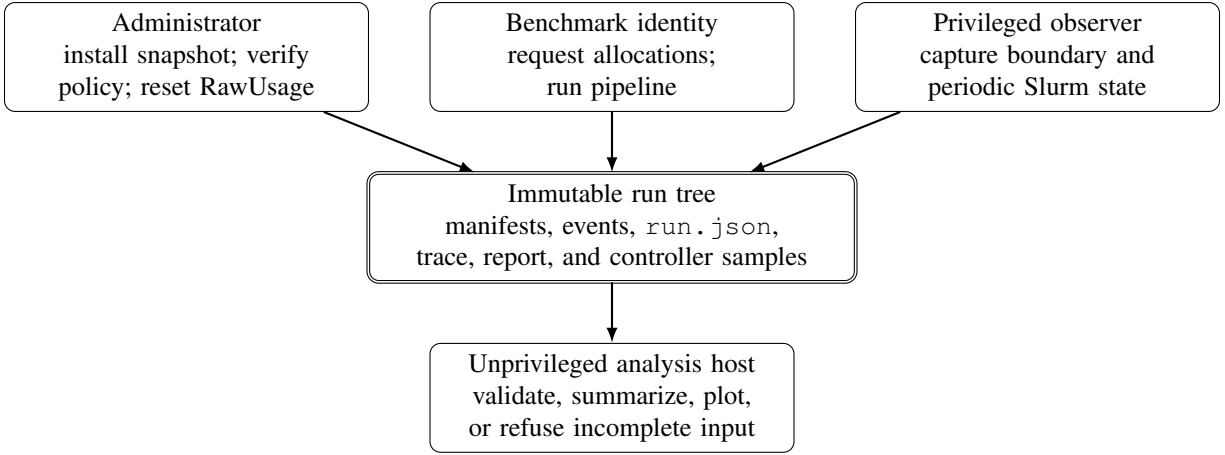
\begin{figure*}[t]
\centering
\begin{tikzpicture}[
  role/.style={draw,rounded corners,align=center,text width=.25\textwidth,
    minimum height=10mm,inner sep=4pt},
  store/.style={draw,double,rounded corners,align=center,text width=.34\textwidth,
    minimum height=10mm,inner sep=4pt},
  flow/.style={-{Latex[length=2mm]},thick},
  node distance=8mm and 8mm]
  \node[role] (admin) {Administrator\\install snapshot; verify policy; reset RawUsage};
  \node[role,right=of admin] (bench) {Benchmark identity\\request allocations; run pipeline};
  \node[role,right=of bench] (observer) {Privileged observer\\capture boundary and periodic Slurm state};
  \node[store,below=of bench] (tree) {Immutable run tree\\manifests, events, \texttt{run.json}, trace, report, and controller samples};
  \node[role,below=of tree] (analysis) {Unprivileged analysis host\\validate, summarize, plot, or refuse incomplete input};
  \draw[flow] (admin) -- (tree);
  \draw[flow] (bench) -- (tree);
  \draw[flow] (observer) -- (tree);
  \draw[flow] (tree) -- (analysis);
\end{tikzpicture}
\caption{Deployment and trust boundary. Privilege is confined to installation,
site-state verification, and the explicitly recorded Fairshare reset. The
benchmark identity generates the measured requests; a separate observer keeps
monitoring traffic out of that identity's per-user RPC delta. Analysis is
offline and requires read access to the immutable run tree, not live Slurm
privilege.}
\label{fig:trust-boundary}
\end{figure*}

The harness makes the experimental clock and attribution rules reproducible in
three parts. Figure~\ref{fig:trust-boundary} shows the role and data boundaries
that keep privileged observation outside the measured identity.

\textbf{Clean-start backend adapters.} Each configuration has a lifecycle
adapter that starts without retained allocations or services, initializes the
strategy, waits for the milestone, validates correctness, and releases every
trial-owned Slurm object. Only adapters are backend-specific.

\textbf{Metric collectors.} Monitoring runs under a separate privileged observer
identity on the scheduler host, outside the measured allocation. Boundary \texttt{sdiag}
snapshots supply the bench user's cumulative RPC count and processing time,
whose boundary differences produce the two attributable RPC measures; a periodic
series records their evolution and cluster context (Table~\ref{tab:factors};
Figure~\ref{fig:timeline}). Observer RPCs are attributed to the observer
identity. Because \texttt{sdiag} itself issues an
RPC~\cite{sdiag_docs}, the
collector limits bench-user monitoring traffic to one \texttt{squeue}
before the pre-boundary, outside the measured window; low-rate \texttt{sdiag}
sampling and drain checks run as the separate observer. Task timing and identifiers come from
the Nextflow trace~\cite{nextflow_trace_docs}; observer-side markers record the
allocation-inclusive start $t^0$ and release boundary. Raw trace, telemetry,
provenance, and correctness artifacts are retained.

\textbf{Reporting.} Offline scripts produce milestone walltimes, per-user RPC
deltas, cluster-wide state, and the trade-off of Section~\ref{sec:results}.

\section{Experimental Design}
\label{sec:design}

\begin{table*}[t]
\caption{Experimental settings and campaign design. Values are frozen per
cluster and treatment as indicated.}
\label{tab:factors}
\centering
\scriptsize
\renewcommand{\arraystretch}{1.0}
\setlength{\tabcolsep}{3pt}
\begin{tabular}{@{}>{\raggedright\arraybackslash}p{0.20\textwidth}
                    >{\raggedright\arraybackslash}p{0.37\textwidth}
                    >{\raggedright\arraybackslash}p{0.37\textwidth}@{}}
\toprule
\textbf{Factor} & \textbf{Dev} & \textbf{Phoenix} \\
\midrule
Deployment backend        & \multicolumn{2}{>{\raggedright\arraybackslash}p{0.746\textwidth}}{four levels (Table~\ref{tab:backends})} \\
\midrule
Software                  & \multicolumn{2}{>{\raggedright\arraybackslash}p{0.746\textwidth}}{Nextflow
                            26.04.0; HyperQueue 0.26.2; Flux 0.88.0;
                            Apptainer 1.4.5-3.el8; container
                            \path{nilablueshirt/make_lastz_chains:latest-amd64}} \\
Workload and request      & \multicolumn{2}{>{\raggedright\arraybackslash}p{0.746\textwidth}}{frozen pipeline,
                            small input producing 4{,}823 LASTZ logical tasks and
                            91 terminal \texttt{CAT\_PSL} tasks; DAG depth 4;
                            production inputs span approximately
                            5{,}000--200{,}000; 1 CPU and 4\,GB per task;
                            15\,min per-task base time limit} \\
Campaign protocol         & \multicolumn{2}{>{\raggedright\arraybackslash}p{0.746\textwidth}}{
                            $N=3$ independent clean-start replicates per backend
                            and cluster (24 accepted runs); randomized sequential
                            backend order; 300\,s observer sampling; 30\,s RPC-free
                            settle tail} \\
Slurm executor settings
                          & \multicolumn{2}{>{\raggedright\arraybackslash}p{0.746\textwidth}}{\texttt{pollInterval=10s};
                            \texttt{queueStatInterval=1min};
                            \texttt{submitRateLimit=100/1min};
                            \texttt{exitReadTimeout=5min};
                            \texttt{killBatchSize=10000}} \\
\midrule
Slurm native/array        & \texttt{queueSize=600}; arrays of 100
                          & \texttt{queueSize=10000}; arrays of 500 \\
Allocation backends       & HQ: partial-node driver allocation and
                            4-worker ceiling (at most 3 exclusive workers
                            concurrent while the driver is active);
                            Flux: \(K=4\) fixed nodes (512 CPU slots)
                          & HQ: demand-driven, at most 300 one-node workers;
                            Flux: \(K=32\) fixed nodes (896 CPU slots) \\
\midrule
Platform                  & \texttt{public}; Slurm 26.05.1; 4 unified nodes;
                            128 CPUs and 512\,GB RAM per node
                          & \texttt{public}; Slurm 25.11.3; 300 unified nodes;
                            28 CPUs and 250\,GB RAM per node \\
Slurm controller limits   & \texttt{MaxArraySize=200000};
                            \texttt{MaxJobCount=400000}
                          & \texttt{MaxArraySize=300000};
                            \texttt{MaxJobCount=550000} \\
Scheduler policy/parameters
                          & \path{preempt_strict_order,nohold_on_prolog_fail,}
                            \path{bf_window=40000,bf_resolution=300,}
                            \path{bf_max_time=60,bf_max_job_user=1000,}
                            \path{bf_max_job_test=100000,bf_continue,}
                            \path{bf_interval=60,max_rpc_cnt=1000}
                          & \path{preempt_strict_order,nohold_on_prolog_fail,}
                            \path{bf_window=43200,bf_resolution=300,}
                            \path{bf_max_time=120,bf_max_job_assoc=2600,}
                            \path{bf_max_job_start=1300,}
                            \path{bf_max_job_test=200000,bf_continue,}
                            \path{bf_interval=30,max_rpc_cnt=450,}
                            \path{enable_user_top} \\
Controlled identity/state & \multicolumn{2}{>{\raggedright\arraybackslash}p{0.746\textwidth}}{bench user,
                            account and QOS fixed per site; placement fixed per backend; RawUsage reset and verified
                            zero} \\
\midrule
Co-tenant background      & none & production co-tenancy via cluster-wide
                            \texttt{sdiag} \\
\bottomrule
\end{tabular}
\end{table*}

\subsection{Backends under test}
\label{sec:design-backends}
The four backends of Table~\ref{tab:backends} run identical pipeline code under
the settings of Table~\ref{tab:factors}. Site-specific operating choices are
frozen separately for each treatment before its first accepted run and draw on prior scheduler-friendly
practice~\cite{mu2026tenpractices}; this paper does not estimate an optimum for
those choices. Because the backends differ in resource ownership and dispatch
locus, the treatments are \emph{deployment strategies} rather than executors
differing in a single implementation detail.

\textbf{Matching rule.} Within a cluster, runs use the same pipeline and input,
bench user, account, QOS, clean-start boundary, and completion gate.
Placement, per-task limits, and allocated concurrency are frozen per treatment
rather than matched: they are part of the deployment strategy.
On Phoenix, native, job-array, and HQ dispatch can reach the partition capacity,
whereas Flux requests a smaller fixed node block (Table~\ref{tab:factors}). The
comparison therefore estimates deployable strategies, not backend effects at
equal concurrency. Each treatment uses Section~\ref{sec:harness}'s clean-start
lifecycle.

Local execution is supported by the harness but excluded because one enclosing
node cannot represent scalable execution across an HPC cluster.

The Flux backend carries one documented fidelity caveat. The
Nextflow Flux executor does not support the per-process \texttt{memory}
directive~\cite{nextflow_executor_docs}, so its runs cannot reproduce the
memory requests that the other backends honor. Slurm instead reserves all memory
on the \(K\) complete nodes for Flux, while Flux does not use the common
per-task memory request for internal admission. The caveat is
reported with the results rather than excluded, because a backend's constraints
are part of what an operator is choosing between.

\subsection{Reference workload and its descriptors}
\label{sec:substrate}
The protocol is not tied to the reference pipeline. Applying it to another
stage-limited Nextflow pipeline requires no process-name allowlist: analysis
infers the terminal process from the latest successful trace completion. We
characterize the workload with descriptors that another site can use for
comparison rather than by name alone. The
class is \emph{wide fan-out}: a small serial preparation stage followed by one
dominant parallel stage of many short, independent tasks. Table~\ref{tab:factors}
reports its cardinality, task request, terminal stage, and DAG depth; dispatch
aggregation is set by the backend (Table~\ref{tab:backends}). These descriptors
define the region of workload space represented here; additional pipelines are
needed to determine how well they predict backend behavior elsewhere.

The instance is \texttt{make\_lastz\_chains}~\cite{kirilenko2023integrating}, a
whole-genome pairwise alignment pipeline built on LASTZ~\cite{harris2007lastz}
and the UCSC Kent toolkit, in the DSL2 form we refactored according to nf-core
conventions in prior work~\cite{mu2025optimizing,mu2026tenpractices}. We use it
as the reference workload because its Fairshare and backfill behavior on a
shared cluster is already characterized, allowing the present study to focus on
the dispatch interface. That prior work reported the native--array walltime
split; new here are the reusable clean-start protocol and bench harness, the
four-strategy comparison, and exact per-user boundary attribution. The milestone
in Table~\ref{tab:factors} isolates the wide fan-out whose dispatch behavior the
study measures. Truncating there keeps the four-backend campaign on two clusters
within a feasible time and resource budget. The stage limit makes LASTZ behave as the
natural end of a single-stage pipeline: every strategy completes the same
scientific work and then performs its normal shutdown.
The trace parser then identifies the corresponding successful task record, so
the milestone is a completed unit of scientific work rather than a log line.

\subsection{Clusters and partitions}
\label{sec:cluster}
We run the campaign on two Slurm clusters with different tenancy models.
Their releases, selected partition geometry, and scheduler policies are
reported in Table~\ref{tab:factors}. Both releases expose the \texttt{sdiag}
per-user RPC table, allowing the collector to extract the dedicated bench
user on each cluster.

\emph{Dev} is a private, four-node, single-user cluster and provides the
low-background reference.
Daemon activity can still appear in its global counters; only the bench
user's per-user delta is directly attributable. Dev's nondefault backfill
settings permit deep queue examination and continued scheduling across lock
yields while trading scheduling frequency and precision for lower
overhead~\cite{slurm_scheduling}.

\emph{Phoenix} is the ASU production cluster used in this study, a
high-throughput machine tuned for workloads of many small tasks, with Slurm,
Fairshare~\cite{yalim2020fairshare}, and the backfill scheduler. Its scheduler is tuned to
admit high per-association backfill volume and elevated RPC concurrency, which
is what lets it absorb WMS fan-out; the relevant
\texttt{SchedulerParameters} and the rationale are documented
in~\cite{mu2026tenpractices,mu2026rmacc}.
Both clusters are operated by ASU Research Computing, which also operates the
separately documented Sol supercomputer~\cite{jennewein2023sol}.

We preserve rather than synchronize these site policies: treatment comparisons
are controlled within each cluster, while Dev and Phoenix are contextual
applications rather than a paired estimate of cluster effects.

Exact
trial intervals are retained in each run's immutable \texttt{run.json},
\texttt{trial.env}, and append-only \texttt{events.tsv}. These artifacts also
record the command-line backend order, bench user, sampling cadence, reset
requirement, and workload/resource hashes.

\subsection{Metrics}
\label{sec:metrics}
Each run reports walltime, attributable per-user RPC demand, cluster-wide
Slurm state, time to the first terminal-process task start, and correctness.
Slurm task control is not measured per run: it follows from the dispatch unit
(Table~\ref{tab:backends}). We define the two primary metrics explicitly.
Our walltime boundary starts at the user action before backend startup or the
first allocation request:

\begin{equation}
T_{b,r}=t^{\mathrm{milestone}}_{b,r}-t^{0}_{b,r}.
\label{eq:clean-time}
\end{equation}

\begin{figure}[t]
\centering
\resizebox{\columnwidth}{!}{%
\begin{tikzpicture}[x=1cm,y=1cm]
\fill[black!8]  (1.2,-0.13) rectangle (3.2,0.13);
\fill[black!22] (3.2,-0.13) rectangle (6.4,0.13);
\fill[black!8]  (6.4,-0.13) rectangle (7.15,0.13);
\fill[black!16] (7.15,-0.13) rectangle (7.8,0.13);
\draw[-{Latex[length=1.6mm]}] (0,0) -- (8.35,0);
\foreach \x/\t in {%
  0.3/{\texttt{sdiag} pre},%
  1.2/{$t^{0}$ (clean start)},%
  3.2/{first terminal-task start},%
  6.4/{$t^{\mathrm{milestone}}$},%
  7.15/{release complete},%
  7.8/{\texttt{sdiag} post}}{%
  \draw[thin] (\x,-0.19) -- (\x,0.19);
  \node[rotate=38,anchor=south west,inner sep=1.5pt,font=\scriptsize] at (\x,0.21) {\t};
}
\draw[decorate,decoration={brace,amplitude=4pt,mirror}] (1.2,-0.30) -- (6.4,-0.30)
  node[midway,below=5pt,font=\scriptsize] {$T_{b,r}$, Eq.~\ref{eq:clean-time}};
\draw[decorate,decoration={brace,amplitude=4pt,mirror}] (0.3,-1.05) -- (7.8,-1.05)
  node[midway,below=5pt,font=\scriptsize] {per-user RPC-demand window};
\end{tikzpicture}%
}
\caption{Measurement boundaries for one trial. The clock opens at $t^{0}$,
before any backend service or allocation request, so backend startup and
allocation acquisition occupy the first light band and count against every
strategy. The dark band is task dispatch and execution. The closing light band
is the correctness gate, main-job exit, and release; the final shaded band is
the fixed RPC-free settle tail. The per-user
RPC window brackets the whole lifecycle: activity after
$t^{\mathrm{milestone}}$ is excluded from walltime but included in RPC demand.}
\label{fig:timeline}
\end{figure}
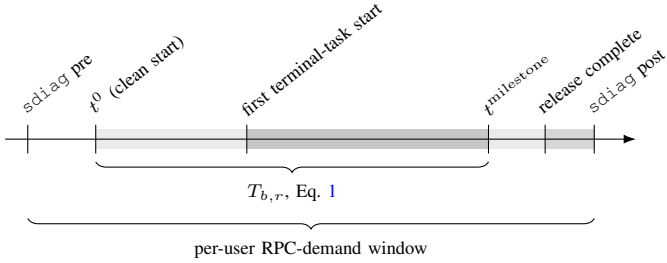

For backend $b$ and replicate $r$, $t^0$ precedes the first backend action and
$t^{\mathrm{milestone}}$ is completion of the final successful task at the common
endpoint. The semantic validator runs after $t^{\mathrm{milestone}}$ and does
not enter $T_{b,r}$; a run enters the comparison only after that gate passes.
The primary RPC-demand metric is normalized request count:

\begin{equation}
C_{b,r}=1000\,
\frac{\Delta\,\mathrm{RPCcount}_{u,b,r}}
     {L^{\mathrm{endpoint}}_{b,r}},
\label{eq:rpc-demand}
\end{equation}

where \texttt{sdiag}'s per-user table supplies the cumulative count for
bench user $u$~\cite{sdiag_docs}, and $L^{\mathrm{endpoint}}$ is the number
of distinct successful logical tasks in the inferred endpoint process. The sensitivity
metric \(P_{b,r}\) replaces \(\Delta\mathrm{RPCcount}\) in Eq.~\ref{eq:rpc-demand}
with cumulative RPC processing time converted from microseconds to seconds.
Figure~\ref{fig:timeline} places both boundaries. The pre-boundary is
taken immediately before $t^0$, and the post-boundary after the fixed
release-settle tail in which the observer issues no RPC\@. Allocation acquisition,
dispatch, and the workflow-release tail are therefore included without repeated
bench-user status polling.
For the fixed input in Table~\ref{tab:factors}, the endpoint cardinality is
constant across runs, so normalization does not change within-cluster ordering;
it also provides a rate for future comparisons across input scales.
The terminal process and task count are derived from each trace rather than
configured as process-name keywords.

RPC count is structural demand attributed to the bench user. Processing
time is controller execution under the conditions present during the run;
contention, queues, and lock availability make it a sensitivity metric rather
than an intrinsic backend property. Table~\ref{tab:results} reports both, while
the periodic samples specified in Table~\ref{tab:factors} provide bench and
non-bench RPCs/min with the observer excluded.

The two primary axes intentionally use different endpoints. Walltime ends at the
scientific milestone, whereas RPC demand continues through correctness checking,
backend teardown, and return to a clean state, capturing the deployment's
complete Slurm-facing lifecycle through the fixed settle tail.

Cluster-wide \texttt{sdiag} readings are secondary context. Cycle-count
differences give main and backfill cycles/min; sampled last-cycle durations,
server threads, and agent queues describe non-attributable system conditions.

\subsection{Sequential execution and cross-run comparability}
\label{sec:design-repl}
The operator supplies a randomized backend order, which the harness executes
sequentially and records. Each backend run is the experimental unit;
samples within its \texttt{sdiag} series are
repeated telemetry, not independent replicates.

The automated protocol accepts any positive number of independent repetitions;
Table~\ref{tab:factors} gives the completed campaign size. We retain individual
observations and report the median with the observed minimum and maximum.
Although the evaluated pipeline is deterministic in its tasks and outputs,
system-level measurements vary across runs. Replication exposes that variability
but supports descriptive comparisons rather than population-level inference.
Native-normalized ratios are paired within replicate and remain descriptive.
The two cluster campaigns may overlap in time, since the
clusters are physically separate; each preserves its rule of one treatment at a
time.

Before every run, the bench user has no pending or running jobs. An
administrator resets that user's RawUsage in the tested account before $t^0$,
verifies the association at zero with \texttt{sshare}, and records the resulting
Fairshare factor. Resource use by the initial wave and its effect on later tasks
are therefore part of the treatment. Sites with contributing parent
associations must instead isolate or match the effective starting
Fairshare~\cite{sshare_docs,sacctmgr_docs}.

Runs are scheduled not to cross \texttt{sdiag}'s default midnight-UTC statistics
reset~\cite{sdiag_docs}.
Any administrative reset or Slurm restart inside a run invalidates its RPC
delta. The scientific task graph is deterministic, and clean-start execution
leaves no service, worker, allocation, or workflow state for the next run. A run
invalidated by infrastructure or execution failure is archived and rerun
independently from a fresh state. The campaign reports only runs that
successfully complete Nextflow and provide a readable trace and report.

\section{Results: Walltime Against RPC Demand}
\label{sec:results}

Every reported strategy must pass the same completion gate
(successful Nextflow completion plus readable trace and report);
a strategy that does
less work is not faster. Table~\ref{tab:results} reports all completed cells as
median [observed minimum--maximum], and Figure~\ref{fig:frontier} retains the
individual observations. The two displayed objectives remain separate because
a site's acceptable RPC demand is a policy choice.

\begin{table*}[t]
\caption{Primary results by cluster. Every cell reports median
[observed minimum--maximum] for $N{=}3$. Walltime includes backend startup and
allocation queueing. RPC-count cells give total count on the first line and
count per 1{,}000 terminal tasks on the second.}
\label{tab:results}
\centering
\footnotesize
\renewcommand{\arraystretch}{1.12}
\setlength{\tabcolsep}{4pt}
\begin{tabular*}{\textwidth}{@{\extracolsep{\fill}}lcccc@{}}
\toprule
\textbf{Backend} & \textbf{Clean-start h} & \textbf{Endpoint-start h} &
\shortstack{\textbf{RPC total}\\\textbf{(/1k)}} & \textbf{RPC s/1k} \\
\midrule
\multicolumn{5}{@{}l}{\emph{Dev (single user)}} \\
Slurm native & 1.01 [0.99--2.18] & 0.99 [0.97--2.17] &
  \shortstack{5{,}120 [5{,}105--5{,}381]\\56{,}264 [56{,}099--59{,}132]} &
  166.74 [166.04--166.94] \\
Slurm job array & 0.81 [0.80--1.47] & 0.79 [0.79--1.46] &
  \shortstack{353 [298--497]\\3{,}879 [3{,}275--5{,}462]} &
  11.25 [10.44--12.63] \\
HyperQueue & 3.82 [3.68--3.84] & 3.82 [3.68--3.83] &
  \shortstack{78 [78--79]\\857 [857--868]} &
  0.55 [0.26--0.68] \\
Flux & 0.81 [0.81--0.84] & 0.80 [0.80--0.83] &
  \shortstack{155 [155--161]\\1{,}703 [1{,}703--1{,}769]} &
  0.22 [0.21--0.23] \\
\midrule
\multicolumn{5}{@{}l}{\emph{Phoenix (shared)}} \\
Slurm native & 1.03 [1.03--1.04] & 1.02 [1.02--1.03] &
  \shortstack{5{,}114 [5{,}111--5{,}126]\\56{,}198 [56{,}165--56{,}330]} &
  7{,}888.45 [6{,}433.49--8{,}008.47] \\
Slurm job array & 0.41 [0.32--0.49] & 0.40 [0.31--0.48] &
  \shortstack{440 [387--545]\\4{,}835 [4{,}253--5{,}989]} &
  532.79 [503.51--834.93] \\
HyperQueue & 0.32 [0.32--0.97] & 0.32 [0.32--0.97] &
  \shortstack{415 [199--419]\\4{,}560 [2{,}187--4{,}604]} &
  425.66 [326.55--448.17] \\
Flux & 0.65 [0.64--0.68] & 0.64 [0.63--0.67] &
  \shortstack{127 [125--131]\\1{,}396 [1{,}374--1{,}440]} &
  99.44 [23.43--103.61] \\
\bottomrule
\end{tabular*}
\end{table*}

\subsection{Walltime and RPC demand}
\label{sec:walltime}

Walltime reports the quantity the researcher experiences. On Phoenix, median
clean-start walltime ordered HyperQueue, arrays, Flux, then native. HyperQueue
had the fastest median, but its lead over arrays was not consistent across
replicates; arrays consistently led Flux. On Dev, array and Flux medians were
nearly equal and their ordering changed by replicate. Both beat native in every
matched run, whereas HyperQueue was consistently slower than native.

The endpoint-start column separates time to the first terminal-process start
from the remainder without pretending that an allocation-backed task and a
Slurm-native task queue in the same place. It shows that nearly all clean-start
walltime elapsed before the short terminal stage. The ordering was therefore
established during resource acquisition and bulk-stage dispatch rather than in
the common terminal process.

Attributable RPC demand had the same broad result on both clusters: every
non-native run issued fewer requests than its paired native run. Flux had the
lowest median demand on Phoenix, whereas HyperQueue had the lowest on Dev.
RPC processing time preserved the Phoenix count ordering but not Dev's: Flux
consumed less processing time there even though HyperQueue issued fewer
requests. Count is therefore the structural metric; processing time remains
scheduler-state-sensitive context.

\subsection{Phoenix cluster-wide Slurm state}
\label{sec:slurmstate}
\begin{figure*}[t]
\centering
\includegraphics[width=0.96\textwidth]{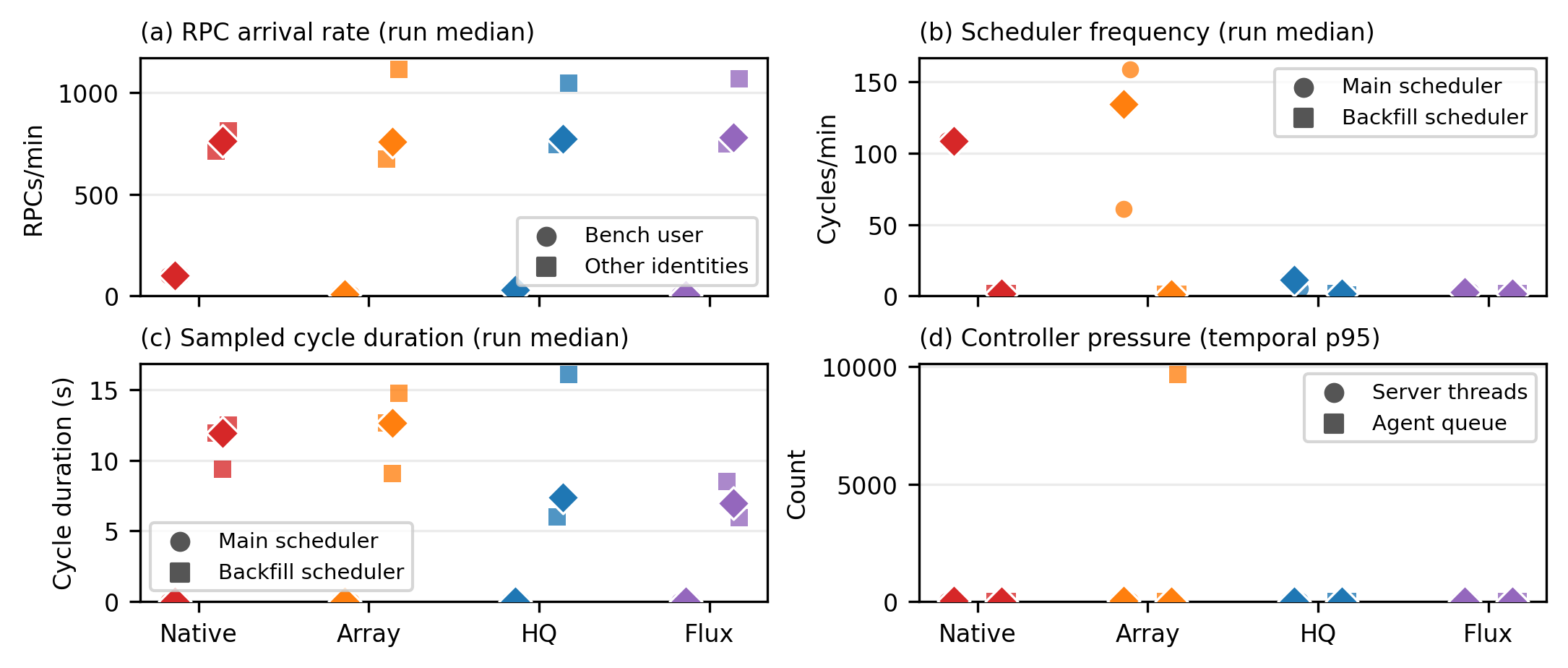}
\caption{Secondary Phoenix system context from periodic samples across the
replicated campaign. Panel (a) separates RPCs attributed to the bench user from other
identities, excluding the observer. Panels (b)--(d) include co-tenant and Slurm
activity and are not attributed solely to a backend. Large diamonds summarize
the run medians; smaller marks retain individual runs.}
\label{fig:sdiag}
\end{figure*}

Figure~\ref{fig:sdiag} shows sustained but varying co-tenant activity that
exceeded the benchmark-user rate throughout the campaign. Native consistently
produced the largest benchmark-user share, while scheduler frequency differed
substantially among treatment windows. An isolated agent-queue excursion also
occurred during an array window. The global scheduler context was not matched
across treatments, and a tested backend can itself affect these metrics; they
are context rather than evidence that background was identical or that a
backend caused a global response.

\subsection{The non-dominated operating points}
\label{sec:frontier}
\begin{figure*}[t]
\centering
\includegraphics[width=0.96\textwidth]{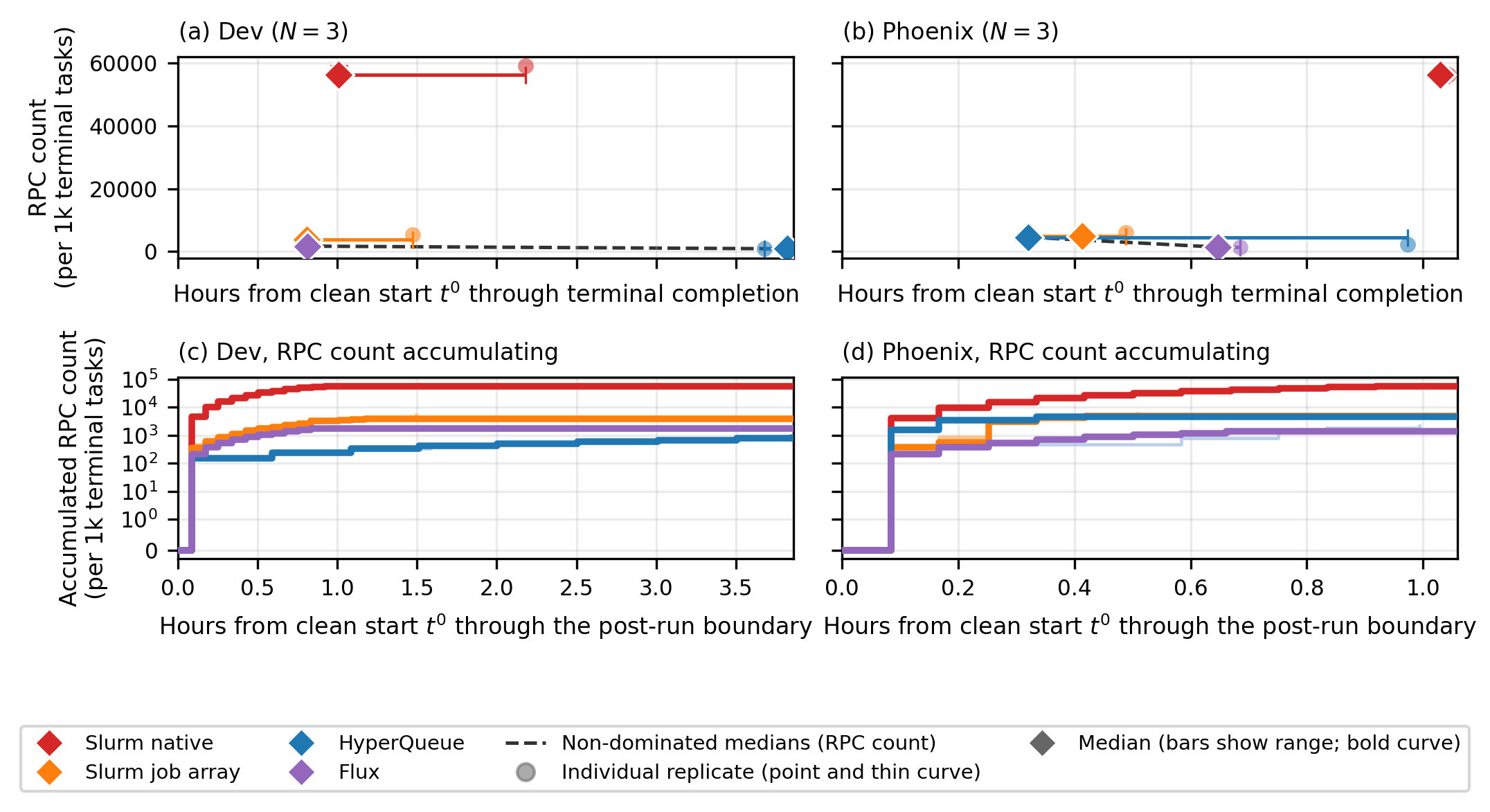}
\caption{Top: clean-start walltime against normalized attributable RPC count
on Dev and Phoenix. Bottom: RPC-count accumulation through the post-run
boundary. Pale marks and curves are individual replicates; diamonds and bold
curves are medians, with bars showing observed ranges. Dashed segments
connect non-dominated median operating points under RPC count. Lower-left is
better.}
\label{fig:frontier}
\end{figure*}

Figure~\ref{fig:frontier} visualizes all completed treatments on each cluster.
Strategy $A$ dominates $B$ only if it is no slower and issues no more
normalized attributable RPCs, with at least one strict
improvement~\cite{deb2002nsga}. Under median walltime and RPC count, the
non-dominated Dev set is arrays, Flux, and HyperQueue; on Phoenix it is
HyperQueue and Flux. Membership is less stable run by run: Flux is the only
strategy that remains non-dominated throughout the Phoenix campaign, while Flux
and HyperQueue do so on Dev. Replacing count with processing time removes
HyperQueue from the Dev median set because Flux is both faster and lower in
processing time. The Phoenix median set remains HyperQueue and Flux.

Points that trade one objective against the other are alternatives for different
site policies, not statistical ties or failures to identify a winner.
Under the matching rule of Section~\ref{sec:design-backends}, allocated
concurrency is an explicit component of each operating point. It would confound
a claim about backend implementation at equal capacity, but that is not the
estimand here: the frontier compares deployable strategies, including the
allocation acquisition time caused by their realistic resource geometries.

The accumulation panels distinguish low sustained request rates from
front-loaded fan-outs while retaining replicate variability. Because the curves
are per-user \texttt{sdiag} deltas rather than trace-derived submissions, they
also cover the two backends that expose no per-task Slurm record. Each curve
ends at the corresponding normalized total in Table~\ref{tab:results}; the RPC
window continues through validation, teardown, and the post-run settle tail.

\section{Operational Interpretation}
\label{sec:tradeoffs}

\subsection{Fairshare, backfill, and task geometry}
\label{sec:tradeoff-fairshare}

Fairshare and RPC aggregation are distinct mechanisms. Fairshare reflects
accumulated TRES usage rather than submission
count~\cite{slurm_priority,sshare_docs}. In prior Phoenix measurements, the
initial LASTZ wave rapidly reduced the bench user's association Fairshare, making
later progress increasingly dependent on
backfill~\cite{mu2025optimizing,mu2026rmacc}. Short, accurate resource requests
therefore remain important regardless of dispatch
backend~\cite{slurm_scheduling}.

Arrays preserve independently schedulable elements while amortizing submission
and polling~\cite{slurm_arrays,nextflow_array_docs}. Task geometry therefore
controls TRES consumption and backfill fit, while array grouping primarily
controls RPC traffic. That separation predicts a cluster-dependent effect, which
is what earlier measurements show: arrays improved walltime on the shared
cluster and not on the single-user one~\cite{mu2026tenpractices}.

\subsection{Acquisition cost versus dispatch relief}
\label{sec:tradeoff-acquisition}

All trials begin from a clean start, so the comparison is between patterns of
resource acquisition rather than between reused and freshly acquired state.
The dispatch units in Table~\ref{tab:backends} imply different acquisition
costs. Table~\ref{tab:results}'s endpoint-start interval captures the combined
time spent acquiring resources and completing earlier pipeline stages, while
total walltime shows whether faster in-allocation dispatch repays those costs by
the milestone. This decomposition matters particularly on a shared partition:
a long $K$-node request may wait longer than many small backfill-friendly
requests even when its measured attributable RPC demand is lower.

\section{Discussion}
\label{sec:discussion}
\label{sec:method-generalization}

\textbf{Applying the method.} A site selects a representative input and
correctness-checked milestone, records Section~\ref{sec:substrate}'s descriptors,
and repeats Eqs.~\ref{eq:clean-time}--\ref{eq:rpc-demand} under fixed policy with
a separate observer. Sites unable to reset an isolated association should block
or match starting Fairshare. They reject strategies violating demand, capacity,
or control constraints, then select the fastest remaining option.
Aggregation hides per-task records, limits, cancellation, and preemption from
Slurm, but the enclosing allocation still accrues aggregate TRES usage
(Table~\ref{tab:backends}).

\textbf{Generalization and threats to validity.} The method targets
standardized Nextflow execution, including nf-core pipelines, on university
Fairshare-based Slurm. The campaign covers one region of the descriptor space,
so it reports the backend ordering observed for this wide fan-out workload,
cardinality, and task geometry rather than every DAG shape.
The protocol applies where Slurm exposes per-user \texttt{sdiag} statistics and
permits a separate observer. Rankings remain conditional on workload, hardware,
policy, tenancy, and background. Replication exposes but does not control the
variation in Phoenix background. Attribution remains per-user, but
global response includes co-tenants. The physically separate clusters are
contextual applications, not a paired estimate of cluster effects. Equal
starting Fairshare does not imply equal priority or ordinary user histories.
The tested input is also substantially smaller than the motivating production
cases summarized in Table~\ref{tab:factors}; total RPC demand and backend
ordering must not be extrapolated to larger cardinalities from this campaign.
Finally,
this study evaluates freshly started services only; persistent HQ or Flux deployments may
occupy different points.

\section{Conclusion and Future Work}
\label{sec:conclusion}

We compared Nextflow deployment strategies using clean-start walltime and
per-user Slurm RPC attribution on private and production clusters. Every
Phoenix alternative reduced walltime and RPC count relative to its paired
native run. Arrays consistently reduced both objectives while preserving
per-element Slurm control; Flux minimized RPC demand; and HyperQueue's fastest
median was not stable across replicates. On Dev, arrays and Flux beat native
walltime in every matched run, while HyperQueue was slowest despite issuing the
fewest requests. The results show
that deployment choice trades allocation acquisition and task-level visibility
against RPC aggregation and workflow turnaround; no strategy is optimal under
every site constraint. The primary contribution is the reusable methodology that produced
the observations rather than a universal ranking. A deployment choice inside a
workflow manager is therefore also a scheduling decision for the site that
hosts it.

Future work will repeat the campaign at several input cardinalities and extend
it to other pipelines and workflow systems, beginning with
Snakemake~\cite{koster2012snakemake} and
Pegasus~\cite{deelman2015pegasus}. It will sweep endpoint cardinality and DAG
shape to test the hypothesis that total RPC demand grows with task count and
that dispatch aggregation becomes increasingly valuable, reporting both total
and per-1,000-task demand. WfBench within
WfCommons~\cite{coleman2022wfcommons} will provide portable synthetic cases.

\section*{Artifact Availability}
The released artifact contains the backend configurations, measurement harness,
raw per-run records, and analysis scripts required to reproduce
Table~\ref{tab:results} and Figures~\ref{fig:sdiag} and~\ref{fig:frontier}, and
supports repetition at arbitrary $N$ without modification.
Source and artifacts are at
\url{https://github.com/NilaBlueshirt/Slurm-Stress-SC26}. The same
repository carries the accepted author version of~\cite{mu2026tenpractices},
whose publication of record is forthcoming.

\section*{Acknowledgments}
We thank the ASU Research Computing team for cluster access and operational
support, the Earth BioGenome Project for the collaboration that motivated the
reference workload, and the upstream \texttt{make\_lastz\_chains} authors at the
Hiller Lab. We thank SchedMD and the HyperQueue and Flux development teams for
their tools.

\IEEEtriggeratref{15}

\end{document}